\documentclass[reqno,11pt]{amsart}

\usepackage{graphicx}
\usepackage{amssymb,amsthm}
\usepackage{amsmath}
\usepackage{amsfonts}
\usepackage{dsfont}

\usepackage[a4paper,  margin=3.6cm]{geometry}

\usepackage[active]{srcltx}
\makeatletter\@addtoreset{equation}{section}\makeatother

\newtheorem{theorem}{Theorem}[section]

\newtheorem{proposition}[theorem]{Proposition}
\newtheorem{remark}[theorem]{Remark}
\numberwithin{equation}{section}

\title[A simple model of tumor growth]{Modeling tumor growth: a simple individual-based model and its analysis}

\author{ Yuri  Kozitsky}

\address{Instytut Matematyki, Uniwersytet Marii Curie-Sk{\l}odowskiej, Plac Marii Curie-Sk{\l}odowskiej 1, 20-031 Lublin, Poland}
\email{jkozi@hektor.umcs.lublin.pl}

\author{Krzysztof Polorz}

\address{Instytut Matematyki, Uniwersytet Marii Curie-Sk{\l}odowskiej, Plac Marii Curie-Sk{\l}odowskiej 1, 20-031 Lublin, Poland}
\email{krzysztof.pilorz@poczta.umcs.lublin.pl}


\begin{document}


\begin{abstract}

Initiation and development of a malignant tumor is a complex
phenomenon that has critical stages determining its long time
behavior. This phenomenon is mathematically described by means of
various models: from simple heuristic models to those employing
stochastic processes. In this chapter, we discuss some aspects of
such modeling by analyzing  a simple individual-based model, in
which tumor cells are presented as point particles drifting in
$\mathbf{R}_{+}:=[0,+\infty)$ towards the origin with unit speed. At
the origin, each of them splits into two new particles that
instantly appear in $\mathbf{R}_{+}$ at random positions. During
their drift the particles are subject to a random death before
splitting. In this model, trait $x\in \mathbf{R}_{+}$ of a given
cell corresponds to time to its division and the death is caused by
therapeutic factors. On its base we demonstrate how to derive a
condition -- involving the therapy related death rate and cell cycle
distribution parameters -- under which the tumor size remains
bounded in time, which practically means combating the disease.
\end{abstract}

\maketitle

\section{Introduction}

Understanding complex systems \index{complex!system} is a paramount
interdisciplinary task of modern science. An efficient way of
achieving this is modeling, which basically assumes elaborating and
studying mathematical objects -- both analytically and numerically.
The following\footnote{
http://www.informatics.indiana.edu/rocha/publications/complex/csm.html}
 typical
`definition' provides the key attributes to such modeling: ``A
complex system \index{complex!system} is any system featuring a
large number of interacting components (agents, processes, etc.)
whose aggregate activity is nonlinear (not derivable from the
summations of the activity of individual components) and typically
exhibits hierarchical self-organization under selective pressures."
The mentioned aggregate activity has -- broadly understood --
critical points in the vicinity of which its character is
drastically different. An instance is provided by the Ising model in
two or more dimensions, see, e.g., \cite{Korder}, the equilibrium
thermodynamic phases of which are multiple for each $T<T_c$, in
contrast to the case of $T>T_c$ where there is only one such phase.
Here $T$ and $T_c$ are the temperature and the critical temperature,
respectively. In each of the multiple phases, there is ordering -- a
nonlinear activity of the kind mentioned above, not derivable from
the individual behavior of single spins. It is absent in the phase
existing at $T>T_c$. This ordering is caused by a spin-spin
interaction, without which nothing like this is possible as there is
only one phase at all $T>0$.

This example from equilibrium statistical physics manifests critical
dependence of equilibrium states of the Ising model on the model
parameters, which is irrelevant to time by the very nature of
equilibrium states. There exists another type of interactions -- and
thus of criticality -- observed in systems that develop in time.
Herein, along with `horizontal' interaction (dependence) between the
constituents existing at a given moment of time, there can be a
`vertical' dependence between states at consecutive time moments. A
significant example here is a system of branching \index{branching}
entities in which each of them splits into some number of new ones.
This number can also be zero meaning the death of the entity. Here
criticality is related to the law of branching, not to interactions
which can be absent at all. In the case of super-critical (resp.
sub-critical) branching the system explodes (resp. dies out) in the
long time limit.

Initiation and progression of a malignant neoplasm is a complex
phenomenon that has critical stages determining its long time
behavior, and hence the outcome of the disease. Its mathematical
modeling is among the most actual problems of applied mathematics.
Being supported by powerful computational means such modeling can
essentially contribute to combating cancer -- one of the most
challenging scientific and social problems of modern life, see
\cite{1,2,Kimm} and the literature quoted therein. Among the
processes to be modeled there is the proliferation of tumor
\index{proliferation!of!tumor} cells subject to therapeutic pressure
caused by chemo- and/or radio-therapy\cite{2,3}. Most of the models
used here are of purely phenomenological nature and operate with
such aggregate parameters as tumor volume or mean number of tumor
cells. They resemble classical thermodynamic models -- predecessors
of microscopic models of statistical physics like the Ising model
mentioned above. In this context,  one might name the
logistic-growth and Gompertz models, as well as their more advanced
versions, see \cite{1}, or those based on taking into account
individual-cell parameters \cite{1,3,4}. Nowadays, it is
well-established that most of the processes in biological tissues --
and, certainly, in malignant neoplasms -- occur at random. This
includes the proliferation of tumor \index{proliferation!of!tumor}
cells by their division, where the language of branching
\index{branching} processes is more than appropriate. With this
regard, we refer the reader to the monograph \cite{KA} where one can
find more on biological aspects of the problem (Chapter 2), as well
as on the mathematical theory of branching (rest of the monograph).

The aim of this chapter is to illustrate the possibilities of the
theory of stochastic branching  phenomena in modeling proliferation
of cancer cells by analyzing a simple individual-based model
proposed recently in \cite{Koz}. This model describes the stochastic
(Markov) dynamics of a population of tumor cells in which every of
its members has programmed division into two new cells after passing
through a cycle of stages. The cycle length  is random. At each
moment of its life, a population member can die before division --
also at random. If it manages to stay alive till the very end of the
cycle -- and thus to produce two progenies -- each of these two
starts its own cycle of random length.  The death rate depends on
the applied therapeutic pressure and is assumed the same for all
cells. Its magnitude that guarantees the extinction of the tumor --
or at least its boundedness in time -- is the key parameter which
the theory has to provide given the distribution of the cell cycle
lengths is known. As we will see below, despite the model
simplicity, it captures the most significant peculiarities of the
stochastic dynamics of populations of cancer cells remaining after
removal of the bulk tumor. Moreover, as a part, this model can be
used in more advanced models which take into account further
peculiarities of the described phenomenon. Note that the study of
this simple model turned to be quite demanding and is based on
rather sophisticated mathematical tools the details of which can be
found in \cite{Koz}.

\section{Beginnings}
In this section, we provide elementary information on the biomedical
aspects of the phenomenon of interest and elementary introduction to
the mathematics related to the model. More details on both these
subjects can be found in \cite{KA,Koz}.
\subsection{Biomedical Aspects}

\begin{figure}
\centerline{\includegraphics[width=8.4cm]{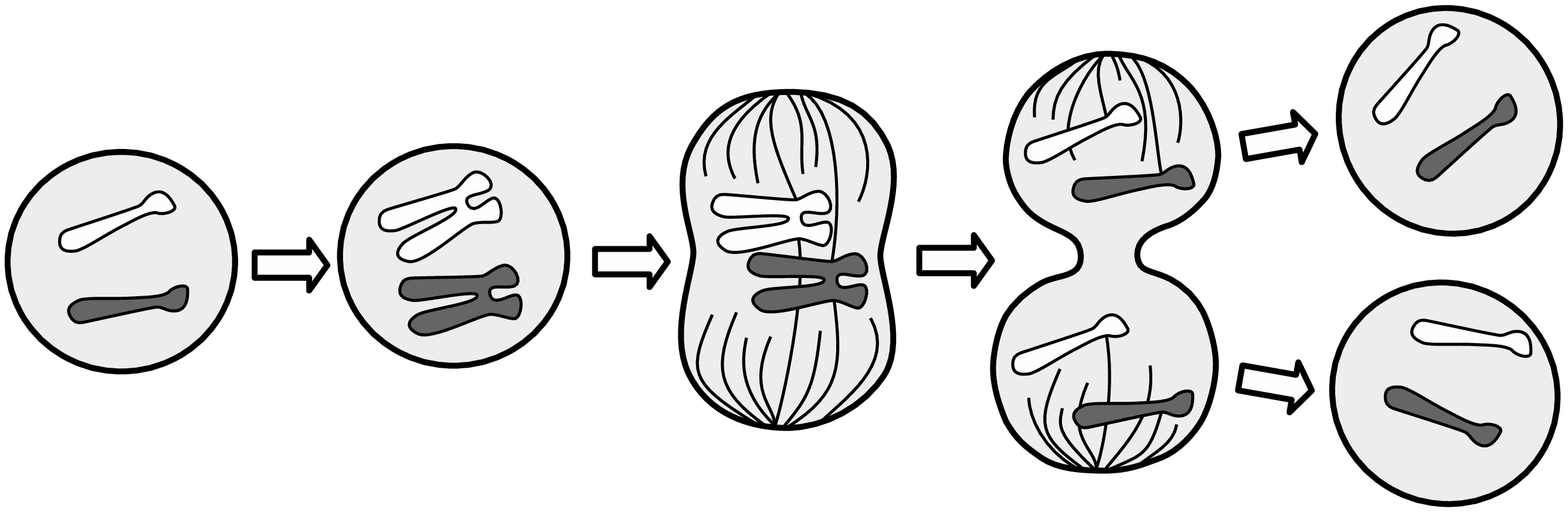}}
\caption{Cell division in eukaryotes.} \label{ka}
\end{figure}

\begin{figure}
\centerline{\includegraphics[width=8.4cm]{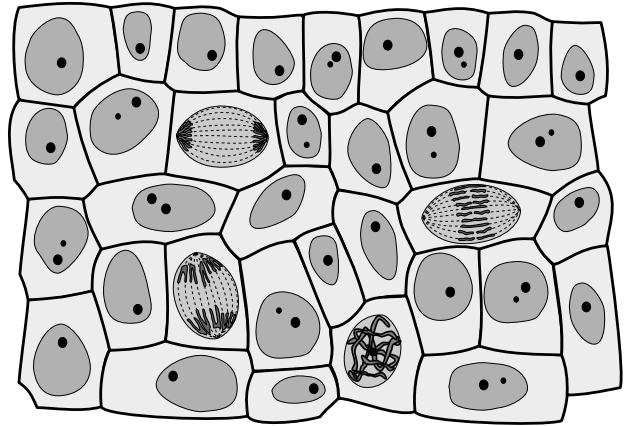}}
\caption{Cell structure of a healthy tissue.} \label{ra_fig1}
\end{figure}

\begin{figure}
\centerline{\includegraphics[width=10.5cm]{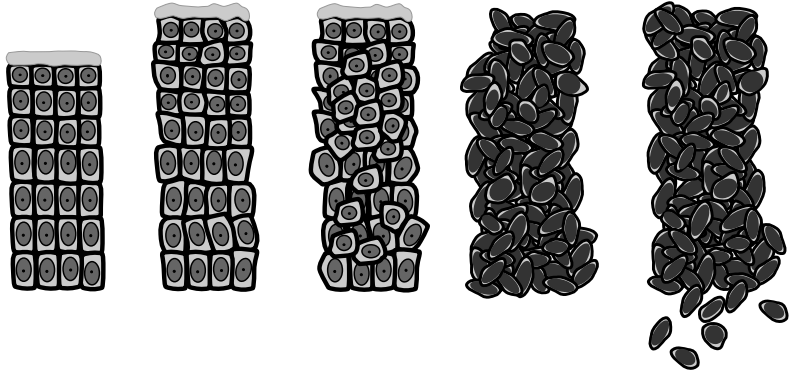}}
\caption{From normal tissue to invasive cancer.} \label{Ig7}
\end{figure}

Each biological substance consists of biological cells that evolve
in time. The only essential evolutionary act of a unicellular
organism is division into two new organisms at the end of the
lifetime interval during which it goes through a sequence of stages,
see Fig. \ref{ka}, including also the DNA replication in the course
of which the genetic information is transferred to the progenies.
It can also happen that a cell dies without division. Due to random
events that occur both in and outside of a cell, its death without
division as well as its lifetime span are random. In multicellular
organisms, their tissues are built up with cells that constitute
quite rigid structure and usually coordinate their evolution with
each other, see Fig. \ref{ra_fig1}.
Normally, developed organisms have more or less constant number of
cells. If a cell dies, its neighbors receive the corresponding
signal, and one of them undergoes division into two new cells. One
of the progenies replaces the died cell, and thereby the overall
balance is restored. There are two types of death: \emph{apoptosis}
and \emph{necrosis}. The first one is a kind of programmed death of
a cell that inevitably occurs to each of them. It is a part of the
mechanism that controls the total number of cells in the organism.
Necrosis is an accidental death that may occur, e.g., due to
external factors. During the division of a cell mutations can occur.
A mutation is the alteration of the nucleotide sequence of the cell
genome. Mostly mutations are irrelevant and the organism functioning
is unchanged. Such mutations are called neutral. Mutations in genes
that regulate cell division, apoptosis, and DNA repair may cause
uncontrolled cell proliferation, in the course of which the total
number of cells gets bigger than usual (hyper- and dysplasia) that
eventually leads to cancer. Such mutations propel the cells
uncontrolled expansion and invasion \cite{5}. "Unlike normal cells,
cancer cells ignore the usual density-dependent inhibition of growth
... piling up until all nutrients are
exhausted\footnote{https://www.biology.iupui.edu/biocourses/N100H/ch8mitosis.html}",
see Fig. \ref{Ig7}, where the pictures going from the left present
normal tissue, hyperplasia, mild dysplasia, severe dysplasia, and
invasive cancer tissue, respectively. This illustrates the way of
initiation of a cancer.

.
\subsection{Branching}

Loosely speaking, branching \index{branching} is a process in which
an entity -- called a particle -- produces at random a random number
of offsprings. They repeat this action after some time. The state
space of the process is the set of nonnegative integers
$\mathbf{N}_0= \{0,1.\dots\}$. One of the simplest examples of
branching is the Galton-Watson process, \index{Galton-Watson} see
Chapter 3 of \cite{KA}. In this case, every particle produces $k\in
\mathbf{N}_0$ offsprings with probability $p_k\geq 0$ --
independently of each other.  The lifetime of all of the particles
is the same. In view of this, one may distinguish generations in
their population. Let $Z_n$ be the number of particles in $n$-th
generation. It is a random variable with values in $\mathbf{N}_0$.
The recurrence between the generations is obviously the following
one
\begin{equation}
  \label{1}
  Z_{n+1} = \sum_{j=1}^{Z_n} X_{n,j}, \qquad n\in \mathbf{N}_0,
\end{equation}
where $X_{n,j}$ is the number of offsprings of $j$-th member of
generation $n$. By our assumption all these random variables
$X_{n,j}$ are independent and identically distributed, and the event
$X_{n,j}=k$ has probability $p_k\geq 0$, where the collection
$(p_k)_{k\in \mathbf{N}_0}$ is assumed given. Note that the sum in
(\ref{1}) has random number of summands, and also that $\sum_k p_k
=1$. Usually, one assumes that the mean number of offsprings is
finite, i.e.,
\begin{equation}
  \label{2}
  a:= \langle X_{n,j} \rangle = \sum_{k=0}^\infty k p_k <\infty.
\end{equation}
If the number of particles in $n$-th generation is known, then the
conditional expected number of them in the next generation is
\[
\langle Z_{n+1} \rangle|_{Z_n}  = \sum_{j=1}^{Z_n} \langle X_{n,j}
\rangle = a Z_n.
\]
By iterating the latter, cf. (\ref{1}), we then get the
unconditional expectation
\begin{equation}
  \label{3}
\langle Z_{n} \rangle = a^n N_0,
\end{equation}
where $N_0$ is the (non-random) number of particles in the initial
generation. The value $a=1$ is \emph{critical}. For $a<1$, the
branching process described by the recurrence (\ref{1}) is
sub-critical, cf. \cite[page 11]{KA}, in which the average number of
offsprings of a particle is less than one. By (\ref{3}) we then get
$\langle Z_{n} \rangle \to 0$ as $n\to +\infty$, that means
extinction of the population. For $a>1$, the branching
\index{branching} process is super-critical, which means that
$Z_n\to +\infty$ with high probability. Note that, for $p_0>0$,
there may exist nonzero probability that the process dies out even
in this supercritical case.
\subsection{Dynamics: deterministic and stochastic}

Now we turn to basic aspects of stochastic evolution. First, we
introduce general notions, and then pay attention to an important
feature of the stochastic counterpart.

\subsubsection{Dynamical systems}

Let $S$ be a nonempty set elements of which are considered as states
of a given system. Such sets are called phase spaces. Usually, $S$
is endowed with mathematical attributes, such as topology and the
corresponding Borel $\sigma$-field of its subsets. This allows one
to define on $S$ probability measures, the set of which is denoted
as $\mathcal{P}(S)$. As an example, one can keep in mind a harmonic
oscillator for which $S= \mathbf{R}^2$ -- the set of pairs $s=
(q,p)$, where real $q$ and $p$ are position and momentum of the
oscillator, respectively. Another example can be $S=\mathbf{N}_0$,
see the Galton-Watson \index{Galton-Watson} model above. In such a
case, in state $n\in \mathbf{N}_0$ the system consists of $n$
elements, say particles. Then a (continuous time) dynamical system
is a map $(t,s) \mapsto s_t\in S$ such that $s_0=s$. Here $t$ is
time and $s$ is the initial state -- origin of the \emph{trajectory}
$(s_t)_{t\geq 0}$. Often, such trajectories are obtained by solving
(if possible) differential equations, called \emph{evolution}
equations. For the mentioned harmonic oscillator, these equations
are
\begin{equation}
  \label{HE}
\dot{q}_t = p_t/m, \qquad \dot{p}_t = - k q_t,
\end{equation}
where dot means time derivative and $m$ and $k$ are oscillator's
mass and rigidity, respectively. Then the  trajectory $(q_t,p_t)$ is
obtained -- as the corresponding trigonometric functions -- by
solving (\ref{HE}). There exists another way of describing such
evolutions, especially useful if the direct solving like in the case
of (\ref{HE}) is impossible. It is based on the use of
\emph{observables}, which are suitable functions $F:S \to
\mathbf{R}$. In this setting, $F(s)$ is the value of observable $F$
in state $s$ and the evolution $F\to F_t$ is defined by the identity
$F_t (s) = F(s_t)$, i.e., it is \emph{backward} in this sense. In
the Hamiltonian case of (\ref{HE}), the backward evolution equation
is
\begin{equation}
  \label{BEE}
  \dot{F}_t(q,p) = \frac{\partial F_t (q,p)}{\partial q} \frac{\partial H (q,p)}{\partial
  p} - \frac{\partial F_t (q,p)}{\partial p} \frac{\partial H (q,p)}{\partial
  q},
\end{equation}
where $H = p^2/2m + k q^2 /2$ is oscillator's Hamiltonian. Equations
like (\ref{HE}), (\ref{BEE}) describe deterministic evolution. To
take into account random events that may occur in the system, one
ought to employ probability measures $\mu\in \mathcal{P}(S)$ as
system \emph{states}. Then the value of observable $F$ in state
$\mu$ is given by the following integral
\begin{equation}
  \label{BE}
  \mu(F) = \int_{S} F(s) \mu(d s),
\end{equation}
with the possibility to include \emph{point states} $s\in S$ into
this picture by associating them with Dirac measures $\delta_s$. The
evolution now is a map $(t, \mu) \mapsto \mu_t\in \mathcal{P}(S)$,
where $\mu$ is the initial state. This evolution is
\emph{deterministic} if $\mu= \delta_s$ implies $\mu_t =
\delta_{s_t}$ for some $s_t\in S$, holding for all $t>0$. In other
words, this evolution preserves the set of Dirac measures.
Otherwise, it is called \emph{stochastic}.

\subsubsection{Honest stochastic evolutions}

Now we turn to the Galton-Watson \index{Galton-Watson} example in
which $S=\mathbf{N}_0$. Let $\mu$ be a state on this $S$. Then it is
defined by its values on singletons $\{n\}$, denoted by $\mu(n)$.
That is, $\mu(n)$ is the probability of the event ``the system
consists of $n$ particles". and the number $\mu(A)$ is the
probability that the state of the system lies in $A\subset S$.
Obviously, $\mu(S)=1$ as $\mu$ is a probability measure. In the
course of evolution $(t,\mu)\mapsto \mu_t$, it may happen that, for
some $t>0$, $\mu_t(S)<1$, i.e., $\mu_t$ fails to satisfy the
mentioned condition, even if the initial state $\mu$ does. In the
mentioned example, this corresponds to
\[
\mu_t(\mathbf{N}_0)= \sum_{n\geq 0} \mu_t(n) <1.
\]
 That is, the
probability of having at time $t$ any finite number of particles is
less than one, and then $1- \mu_t(\mathbf{N}_0)>0$ is the
probability that the system is infinite at this time, which means
its explosion. Thus, the system explodes with positive probability
if this occurs. This is similar to the extinction with positive
probability of a supercritical branching \index{branching} process
mentioned above. The evolution $(t,\mu)\mapsto \mu_t$ such that
$\mu_t(S)=1$ for all $t>0$ is called \emph{honest}. In this case, no
explosion occurs. Obviously, honesty of the evolution of population
of tumor cells is an extremely important aspect of the theory.
Further details on honest stochastic evolutions can be found
\cite{hon,Mustapha,Reuter}.

\section{The Model}

As mentioned above, we are aiming at showing the power of modeling
with the help  of an individual-based model\cite{Koz}  that
describes the proliferation of tumor cells. Here ``individual-based"
means that the evolution of each single cell is taken into account
explicitly -- in contrast to phenomenological models
\cite{1,5,LR,Rot} where a population of cells is considered as a
medium characterized by, e.g., density. Before introducing the
model, we formulate basic principles and provide heuristic arguments
intimating possible outcomes of its study.

\subsection{Basic arguments}

\begin{figure}
\centerline{\includegraphics[width=11cm]{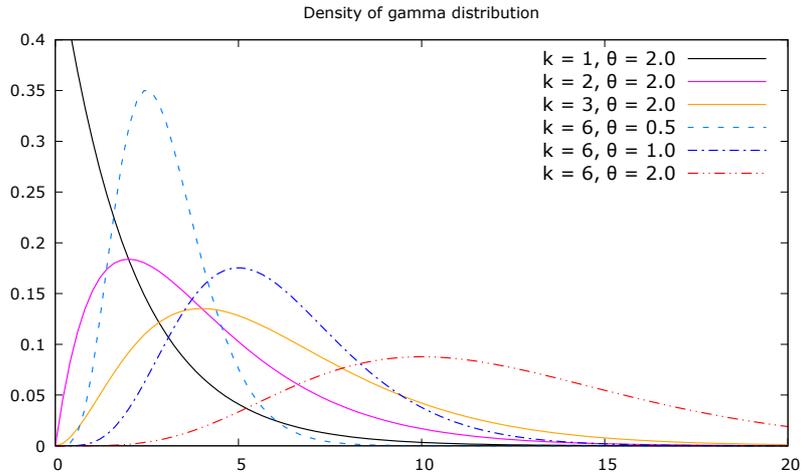}} \caption{Density
$g$ of the $\varGamma$-distributions, see (\ref{Gd}), for various
values of $k$ and $\theta$.} \label{Ig8}
\end{figure}

A standard approach to curing cancer can schematically be presented
as follows. The main part of the bulk tumor is removed by surgery,
and the remaining tumor cells are then treated by chemo- and/or
radio-therapy aiming at their extinction. As the therapy can also
affect healthy tissue, an essential aspect of the method is
minimizing the therapeutic pressure needed to achieve the aim. The
considered model is intended to describe the evolution of the
remaining population of tumor cells and thus to estimate their
minimal mortality that guarantees the mentioned extinction. Its
construction is based on the following principles.
\begin{itemize}
  \item[(a)] The population of cells is finite. Each of its members
  is characterized by its lifetime (length of its cycle).
  The lifetimes of the cells are independent and identically distributed
  random variables, the common distribution of which is known. At
  the end of its cycle, a cell divides into two progenies.
\item[(b)] Each cell can die before producing
progenies. The death is caused solely by the therapeutic pressure
and is independent of the total number of cells. That is, we do not
take into account natural death (untreated tumor cells are
`immortal', cf. \cite[page 28]{KA}) and competition-caused mortality
-- essential in the bulk tumor and minor after its removal.
\end{itemize}
According to (a), the lifetime $\tau$ of a given cell is random.
Assume for a while that it is deterministic and the same for all
cells, that is the cells behave as in the Galton-Watson
\index{Galton!Watson} model mentioned above, with strictly positive
$p_0$ and $p_2$ and $p_k=0$ otherwise. Clearly, $p_0$ is the
probability of the premature death (due to therapy), and $p_2 =
1-p_0$. To calculate $p_0$ we need to choose the way of realizing
the therapeutic pressure. In its simplest and most realistic
version, the probability of staying alive for a given cell
diminishes with constant speed $- m$, where the mortality parameter
$m>0$ is assumed to be the same for all cells. Its value depends
only on the therapy and (in principle) may be estimated, e.g., in
vitro. According to this, the probability in question is $e^{-mt}$.
At the end of the life period we have $t=\tau$; hence, $p_2 = e^{-m
\tau}$ and $p_0 = 1 - e^{-m \tau}$. Then the branching parameter $a$
is, cf. (\ref{2}),
\begin{equation}
  \label{4}
 a (\tau) = 0\cdot (1 - e^{-m \tau}) + 2 \cdot e^{-m \tau} = 2  e^{-m
 \tau}.
\end{equation}
Now we take into account that $\tau$ is random. Assume that its
probability distribution has density (with respect to Lebesgue's
measure) given by an appropriate function $g$. Then the averaged
branching parameter is
\begin{equation}
  \label{5}
\langle a \rangle = \int_{0}^{+\infty} a (\tau) g (\tau) d \tau  =
2\int_{0}^{+\infty} e^{-m\tau} g (\tau) d \tau= 2 \widehat{g}(m),
\end{equation}
where $\widehat{g}$ is the Laplace transform of $g$, see, e.g.,
\cite{Lap}. Now the extinction condition takes the form
\begin{equation}
  \label{6}
  \widehat{g}(m) < \frac{1}{2}.
\end{equation}
Since $g$ is positive and integrable, $\widehat{g}(m)$ decays to
zero in a monotone way as $m\to +\infty$. At the same time,
$\widehat{g}(0)=1$ due to normalization. Thus, (\ref{6}) can be
satisfied at the cost of large enough mortality. Let $m_*$ be the
(unique) solution of the equation
\begin{equation}
  \label{6b}
\widehat{g}(m) = \frac{1}{2}.
\end{equation}
Then (\ref{6}) is satisfied for all $m>m_*$. For various kinds of
tumor, the distribution of $\tau$ is well-studied, see
\cite{Kimm,Dolb} and also \cite{Gab,Tys,Yates}. Usually, one takes
\begin{equation}
  \label{Gd}
g(\tau) = \frac{\tau^{k-1} e^{-\tau/\theta}}{\theta^k \varGamma(k)},
\qquad k, \theta \in (0,+\infty),
\end{equation}
that is the density of the $\varGamma$-distribution, cf.
\cite{Gab,Yates} and see Fig. \ref{Ig8}. Here $\varGamma (k)$ is
Euler's $\varGamma$-function. \index{$\vbarGamma$-distribution}
In this case,
\begin{eqnarray}
  \label{Gamma}
\widehat{g}(m) & = & \frac{1}{\theta^k \varGamma(k)}
\int_{0}^{+\infty} \tau^{k-1} \exp\left( - m \tau  -
\frac{\tau}{\theta}\right) d \tau \\[.2cm] \nonumber & = & \frac{1}{(1+m\theta)^k
\varGamma(k)} \int_{0}^{+\infty} x^{k-1} e ^{-x} d x  \\[.2cm] \nonumber & = &
(1+m\theta)^{-k},
\end{eqnarray}
and then the condition in (\ref{6}) is satisfied for $m>m_*$ where
\begin{equation}
  \label{6a}
  m_* = \frac{1}{\theta} \left( 2^{1/k} - 1\right).
\end{equation}

\subsection{Towards introducing the model}

In accordance with the principles formulated above, the model in
words can be described as follows. Consider a finite subset of
$\mathbf{R}_{+}:= [0,+\infty)$ -- a \emph{cloud} of point particles.
Then the coordinate $x\in \mathbf{R}_{+}$  of a particle in this
cloud is considered as its time to division. The basic act of the
evolution is \emph{aging} \index{aging} -- drifting towards the
point $x=0$ with unit speed. That is, we assume time to division
diminishes with speed one. By reaching the origin the particle
divides into two new particles -- progenies -- that appear at random
positions on the half-line $\mathbf{R}_{+}$. Thereafter, the
progenies start drifting towards $x=0$. During its lifetime, i.e.,
before division, each particle can be removed at random with
constant (mortality) rate $m>0$. Since the point states of the
system are mentioned clouds, to describe them we will employ notions
and methods of the theory of point processes \cite{DV}.
\index{point!process}

Let $\Gamma$ denote the set of all finite subset of
$\mathbf{R}_{+}$. Its elements are finite clouds mentioned above.
This is the phase space of the population of tumor cells for our
model. It is equipped with the weak topology \index{weak!topology}
which is metrizable in such a way that the corresponding metric
space is separable and complete. Note that a complete
characterization of the weak topology is: a sequence,
$\{\gamma_n\}_{n\in \mathbf{N}}\subset \Gamma$, is convergent in
this topology to some $\gamma\in \Gamma$ if
\[
 \sum_{x\in \gamma_n} g(x) \to \sum_{x\in \gamma} g(x),
\]
that holds for all bounded continuous functions $g:\mathbf{R}_{+}
\to \mathbf{R}$. Let $\mathcal{B}(\Gamma)$ be the corresponding
Borel $\sigma$-field. A function, $f:\Gamma \to \mathbf{R}$, is then
measurable if there exists a collection of symmetric Borel functions
$f^{(n)}: \mathbf{R}^n_{+} \to \mathbf{R}$, $n\in \mathbf{N}$, such
that
\begin{equation}
 \label{0}
f(\{x_1, \dots , x_n\}) = f^{(n)} (x_1, \dots , x_n), \qquad n\in
\mathbf{N}.
 \end{equation}
We also set $f^{(0)} = f(\varnothing)$. In expressions like $\gamma
\cup x$, $x\in \mathbf{R}_{+}$ we consider $x$ as a single-element
configuration $\{x\}$. The Lebesgue-Poisson measure $\lambda$ on
$(\Gamma, \mathcal{B}(\Gamma))$ is defined by the integrals
\begin{eqnarray}
  \label{L1}
&  &  \int_{\Gamma} f(\gamma) \lambda (d \gamma) = f^{(0)} +
  \sum_{n=1}^\infty \frac{1}{n!} \int_{\mathbf{R}_{+}^n}
  f^{(n)} (x_1 , \dots , x_n) d x_1 \cdots  d x_n ,
\end{eqnarray}
holding for all bounded measurable $f:\Gamma\to \mathbf{R}$. Such
integrals have the following evident property
\begin{equation}
  \label{La}
  \int_{\Gamma} \left(\sum_{\xi \subset \gamma} f(\gamma, \xi)
  \right) \lambda ( d \gamma) = \int_{\Gamma}\int_{\Gamma} f
  (\gamma\cup \xi, \xi) \lambda ( d \gamma) \lambda ( d \xi).
\end{equation}
Let $\mathcal{E}$ denote the real Banach space $L^1(\Gamma, d
\lambda)$. Its positive elements constitute the cone
$\mathcal{E}^{+}$. The norm of $\mathcal{E}$ then is
\begin{equation}
  \label{norm}
  \| f \| = \int_{\Gamma} |f(\gamma)| \lambda ( d \gamma).
\end{equation}
Hence, probability densities are elements of $\mathcal{E}^{+}$ of
unit norm.

For a given $n\in \mathbf{N}$, by $\mathcal{W}^{1,1}_n$ we denote
the standard Sobolev space\cite{Maz} on $(0,+\infty)^n$, whereas
$\mathcal{W}^{1,1}_{n,s}$ will stand for its subset consisting of
all symmetric $u$, i.e., such that $u(x_1 , \dots , x_n) =
u(x_{\sigma(1)}, \dots x_{\sigma(n)})$ holding for all permutations
$\sigma \in \varSigma_n$.
\begin{remark}
 \label{1rk}
By Theorem 1, page 4 of Ref. \cite{Maz} we know that each element of
$\mathcal{W}^{1,1}_{n,s}$ -- as an equivalence class -- contains a
unique (symmetric) $u:\mathbf{R}^n_{+} \to \mathbf{R}$ such that
\begin{itemize}
\item[(a)] for Lebesgue-almost all $(x_1, \dots, x_{n-1})$, the map
$\mathbf{R}_{+}\ni y \mapsto u(y, x_1 \dots , x_{n-1})$ is
continuous and its restriction to $(0,+\infty)$ is absolutely
continuous;
\item[(b)] the following holds
\begin{equation*}
  \int_{\mathbf{R}_{+}^n}\left| \frac{\partial}{\partial x_1} u(x_1 , \dots , x_n)\right| dx_1 \cdots d
  x_n < \infty.
\end{equation*}
\end{itemize}
In the sequel, we will mean this function $u$ when speaking of a
given element of $\mathcal{W}^{1,1}_{n, s}$.
\end{remark}
Let $f$ and $f^{(n)}$ be as in (\ref{0}), (\ref{L1}). Let also
$\mathcal{W}$ be then the  set of all $f$ for which
$f^{(n)}\in\mathcal{W}^{1,1}_{n, s}$. Define
\begin{eqnarray}
  \label{11}
  (D f)^{(n)}(x_1, \dots , x_n) & = & \sum_{j=1}^n
  \frac{\partial}{\partial x_j} f^{(n)} (x_1, \dots , x_n) \\[.2cm]
  & = & \frac{d}{dt} f^{(n)} (x_1+t , \dots , x_n+t )|_{t=0}.
  \nonumber
\end{eqnarray}
This allows us to define also
\begin{gather}
  \label{12a}
  (Df)(\gamma) =  \frac{d}{dt} f(\gamma_t)|_{t=0}, \\[.2cm] \nonumber
  f(\gamma_t)  =  f(\gamma) + \int_0^t (D f) (\gamma_\tau)  d\tau.
\end{gather}
Note that \begin{eqnarray}
  \label{12}
  \| D f\|&:=& \sum_{n=1}^\infty \frac{1}{n!} \int_{\mathbf{R}_{+}^n} \sum_{j=1}^n \left| \frac{\partial}{\partial x_j}
  f^{(n)}
  (x_1 ,\dots , x_n) \right| d x_1 \cdots dx_n \\[.2cm] \nonumber & = &  \sum_{n=0}^\infty \frac{1}{n!} \int_{\mathbf{R}_{+}^{n+1}} \left| \frac{\partial}{\partial x}
  f^{(n+1)}
  (x, x_1 ,\dots , x_n) \right| d x d x_1 \cdots dx_n <\infty,
\end{eqnarray}
whenever $f\in \mathcal{W}$. The key issue in (\ref{12}) is the
convergence of the series. In (\ref{12a}), we use shifts of
$\gamma\in \Gamma$. For $t\in \mathbf{R}$, we set $\gamma_t= \{ x+t:
x\in \gamma\}$. For $t>0$, this is well-defined for all $\gamma$,
whereas for $t<0$ one should apply such shifts only to proper
$\gamma$, i.e., such that $x+t \geq 0$ for all $x\in \gamma$.

For $f\in \mathcal{W}$, we define
\begin{equation}
 \label{12A}
\|f\|_{\mathcal{W}} = \|f \| + \|Df\|,
\end{equation}
which  is finite in view of (\ref{12}), see also (\ref{norm}). It is
possible to prove the following statement, see Proposition 2.2 in
\cite{Koz}.
\begin{proposition}
  \label{1pn}
The set $\mathcal{W}$ equipped with the norm defined in (\ref{12A})
is a Banach space. Thus, the linear operator $(D, \mathcal{W})$
defined on $\mathcal{E}$ in
  (\ref{11}) and (\ref{12}) is closed.
\end{proposition}
\

\subsection{Defining the model}

In our approach, the stochastic evolution of the considered
population of tumor cells is Markovian. According to the basic
principles formulated above it is described by the following
backward Kolmogorov equation
\begin{equation}
  \label{M4}
  \frac{d}{dt}F_t = L^* F_t, \qquad F_t|_{t=0} = F_0,
\end{equation}
with
\begin{eqnarray}
  \label{M5}
  ( L^* F)(\gamma)& = & - ( D F)(\gamma) + \sum_{x\in \gamma} m
  \left[ F(\gamma\setminus x) - F(\gamma)\right]\\[.2cm] \nonumber &
  + & \sum_{x\in \gamma} \delta (x) \int_{\mathbf{R}_{+}^2} G(y,z) \left[
  F(\gamma \setminus x\cup \{y,z\}) - F(\gamma)
  \right] dy d z,
\end{eqnarray}
where $F_t:\Gamma \to \mathbf{R}$ is an observable. Here (\ref{M4})
(with $L^*$ given in (\ref{M5})) corresponds to the backward
evolution equation (\ref{BEE}) mentioned above. The first term in
(\ref{M5}) describes aging -- the drift of the trait ``time to
division" towards the origin, and thus is of gradient form. The
second term describes the mortality caused by the therapy with
mortality rate $m\geq 0$ -- the same as in (\ref{4}), (\ref{5}). The
last term describes the division of the cells. Therein, $\delta(x)$
is the Dirac $\delta$-function and $G(x,y)$ is the probability
density of the distribution of lifetimes of the two progenies. It
thus satisfies the normalization condition
\begin{equation*}
\int_{\mathbf{R}_{+}^2} G(x,y) d x d y =1.
\end{equation*}
Obviously, $G(x,y)$ is symmetric and such that
\begin{equation*}
  g(x) = \int_{\mathbf{R}_{+}} G(x,y) dy,
\end{equation*}
is the same as in (\ref{4}) and (\ref{5}). Since the right-hand side
of (\ref{M5}) contains a distribution, possible solutions of
(\ref{M4}) ought to be distributions as well, which may cause
essential technical problems. To avoid them one can pass to a
forward Kolmogorov equation, called also Fokker-Planck equation.
\index{Fokker-Planck!equation} To this end one employs the identity
\begin{equation}
\label{Rule}
 \int_{\Gamma} F(\gamma) (Lf)(\gamma) \lambda ( d \gamma) =
\int_{\Gamma} (L^*F)(\gamma) f(\gamma) \lambda ( d \gamma),
\end{equation}
and the rule (\ref{La}). After some calculations it yields the
Fokker-Planck equation \index{Fokker-Planck!equation}
\begin{equation}
  \label{Ma}
  \frac{d}{dt} f_t = L f_t, \qquad f_{t}|_{t=0} = f_0.
\end{equation}
Here  $f_0$ is the probability density of the initial state $\mu_0$.
That is, $$\mu_0 ( d \gamma) = f_0(\gamma) \lambda ( d \gamma),$$
where $\lambda$ is the Lebesgue-Poisson measure defined in
(\ref{L1}). Likewise, $f_t$ is the probability density of the state
$\mu_t$ at time $t>0$. In accordance with (\ref{Rule}), the operator
in (\ref{Ma}) has the following form
\begin{gather}
  \label{M2}
(Lf)(\gamma) = (Df)(\gamma) + m \int_{\mathbf{R}_{+}} f
(\gamma\cup x) dx \\[.2cm] \nonumber - m|\gamma| f(\gamma) + 2\sum_{\{x,y\}\subset
\gamma} G(x,y) f (\gamma \setminus \{x,y\} \cup 0),
\end{gather}
where $|\gamma|$ is the number of points in $\gamma\in \Gamma$. Our
aim is to solve (\ref{Ma}) and thereby to describe the evolution of
the population.
 Regarding $G$
we will assume the following. For $\beta >0$, we define
\begin{equation*}
  \psi_\beta (x) = \frac{1}{(1+x)^\beta}, \qquad x\geq 0.
\end{equation*}
Then the cell cycle distribution is such that the probability
density $G$ satisfies the condition: there exist $b>0$ and $\beta
\geq 3$ such that, for all $x,y \geq 0$, the following holds
\begin{equation}
  \label{norm4}
  G(x,y) \leq b\left[ \psi_{\beta +1}(x) \psi_\beta (y) + \psi_{\beta}(x) \psi_{\beta+1} (y)
  \right].
\end{equation}
The equation in (\ref{Ma}) should be considered in the Banach space
$\mathcal{E}$ introduced above. The usual way of studying such
evolution equations is to use strongly continuous semigroups of
bounded linear operators in such spaces\cite{Banasiak,Pazy,TV}. To
this end, one has to define $L$ as an unbounded linear operator in
$\mathcal{E}$, which includes also defining its domain. We begin
this by writing
\begin{gather}
  \label{M6}
L = A + B = A + B_1 + B_2, \\[.2cm] \nonumber
(Af)(\gamma) = ( Df)(\gamma) - m|\gamma| f(\gamma), \\[.2cm]
\nonumber (B_1f)(\gamma) = 2 \sum_{\{x,y\}\subset \gamma} G(x,y)
f(\gamma
\setminus \{x,y\}\cup 0), \\[.2cm]
\nonumber (B_2f)(\gamma) = m\int_{\mathbf{R}_{+}}  f (\gamma \cup x)
d x.
\end{gather}
Note that both $B_i$ are positive. Without treatment tumor cells
would certainly proliferate ad infinitum. In view of this, from now
on we  assume that the mortality rate $m$ is strictly positive, and
then  set
\begin{equation}
  \label{M9}
  h_m(\gamma) = 1 + m|\gamma|.
\end{equation}
Recall that $|\gamma|$ denotes the number of points in $\gamma$.
Along with the space $\mathcal{E}$ we also use the following
weighted Banach space $\mathcal{E}_m = L^1(\Gamma, h_m d \lambda)$
equipped with the norm
\begin{equation}
  \label{norm8}
  \|f\|_m = \int_{\Gamma} |f(\gamma) | h_m (\gamma) \lambda (
  d\gamma).
\end{equation}
By (\ref{norm8}) and then by (\ref{M6}), (\ref{M9}) one gets
\begin{equation}
  \label{M12}
  \|B_2 f\| \leq \|f\|_{m}.
\end{equation}
For positive $f\in \mathcal{W}$, by means of (\ref{La}) one can
produce the following calculations
\begin{eqnarray}
  \label{M11}
\|B_1 f \| & = & 2 \int_{\Gamma}\left( \sum_{\{x,y\} \subset \gamma}
G(x,y)
f(\gamma\setminus \{x,y\}\cup 0)\right) \lambda (d \gamma) \\[.2cm] \nonumber & = &
\int_{\Gamma}\left( \sum_{x\in \gamma} \sum_{y\in \gamma\setminus x}
G(x,y) f(\gamma\setminus \{x,y\}\cup 0)\right) \lambda (d \gamma)
 \\[.2cm] \nonumber & = & \int_{\Gamma} \left(
\int_{\mathbf{R}{+}} \sum_{y\in \gamma} G(x,y) f(\gamma\setminus
y\cup 0) d x \right) \lambda ( d
\gamma) \\[.2cm] \nonumber & = & \int_{\Gamma}    \left(
\int_{\mathbf{R}_{+}^2} G(x,y) d x dy\right) f(\gamma\cup 0) \lambda
( d \gamma) = \|D f\|.
\end{eqnarray}
That is, $B_1$ and $D$ can be defined on $\mathcal{W}\subset
\mathcal{E}$. Keeping this and (\ref{M12}) in mind we set
\begin{equation}
  \label{M13}
\mathcal{D}(A)= \mathcal{W}\cap   \mathcal{E}_{m}, \qquad
\mathcal{D}^{+}(A)= \mathcal{D}(A) \cap \mathcal{E}^{+}.
\end{equation}
By (\ref{M12}) and (\ref{M11}) we then conclude that
\begin{equation*}
  B: \mathcal{D}(A) \to \mathcal{E}.
\end{equation*}

\subsection{The result}

Along with $h_m$ defined in (\ref{M9}) we use
\begin{equation}
  \label{M27}
  w_{\sigma,\alpha} (\gamma) = 1 +\sigma |\gamma| + \sum_{x\in \gamma}
  e^{-\alpha x}, \qquad \gamma \in \Gamma,
\end{equation}
with some positive $\sigma$ and $\alpha$. Define, cf. (\ref{norm8}),
\begin{equation}
  \label{norm20}
  \|f\|_{\alpha,\sigma} = \int_{\Gamma} |f(\gamma)| w_{\sigma,\alpha}
  (\gamma) \lambda ( d \gamma).
\end{equation}
 Recall that we assume (\ref{norm4}) holding with $\beta
\geq 3$ and $b>0$. Keeping this in mind we then set
\begin{equation}
  \label{M27z}
  m_0 = \left\{ \begin{array}{ll} \frac{(\beta -1)(b- \beta)}{2 \beta - 5}, \qquad \ \
  &{\rm if} \ \ b>\beta;\\[.4cm]
0 \qquad &{\rm otherwise}.
  \end{array} \right.
\end{equation}
Along with this parameter we also introduce
\begin{equation}
  \label{6c}
  m_1 = \max\{m_0; m_*\},
\end{equation}
where $m_*$ is defined in (\ref{6b}). In the case of
$\varGamma$-distributions, it is given in (\ref{Gamma}) and
(\ref{6a}).

Let us now make precise in which sense we are  going to solve the
Cauchy problem in (\ref{Ma}). By its classical solution, cf. Chapter
4 in \cite{Pazy}, with $f_0 \in \mathcal{D}(L)$ we understand a
function $t\mapsto f_t\in \overline{\mathcal{D}(A)}\subset
\mathcal{E}$ which is: (a) continuously differentiable at all $t\geq
0$; (b) such that both equalities in (\ref{Ma}) are satisfied. Here
$\overline{\mathcal{D}(A)}$ denotes the domain of the closure of
$L=A+B$, which is the closure of $\mathcal{D}(A)$ in the graph norm.
Then the main statement describing the evolution of the considered
population of tumor cells reads as follows, see Theorem 2.7 in
\cite{Koz}.
\begin{theorem}
  \label{1tm}
Assume that (\ref{norm4}) holds with some $\beta\geq 3$ and $b>0$.
Then, for each $m> m_0$ (defined in (\ref{M27z})) and $f_0\in
\mathcal{D}^{+}_1 (A) := \{ f\in \mathcal{D}^{+}(A): \|f\|=1\}$, see
(\ref{M13}), the Fokker-Planck equation (\ref{Ma}) has a unique
classical positive solution $f_t$ such that $\|f_t\|=1$.
Furthermore, for $m>m_1$ defined in (\ref{6c}), there exists $\sigma
>0$ for which $\|f_t\|_{\alpha,\sigma} \leq \|f_0\|_{\alpha,\sigma}$
holding for all $t>0$.
\end{theorem}
The meaning of this mathematical statement will be discussed in the
concluding part of the chapter.

\subsection{Sketch of the proof}

The proof of Theorem \ref{1tm} is based on a version of the
perturbation theory for generators of \index{stochastic!semigroup}
stochastic semigroups \cite{TV}. Its details are similar to those of
the proof of the corresponding statement in \cite{Koz}. Here we just
outline its main steps. One begins by proving that, for each
$\varepsilon
>0$, the operator
\[
L_\varepsilon := A + \varepsilon B,
\]
see (\ref{M6}), with domain $\mathcal{D}(L_\varepsilon) =
\mathcal{D}(L)$ defined in (\ref{M13}), is the generator of a
substochastic semigroup $S_\varepsilon = \{S_\varepsilon
(t)\}_{t\geq 0}$. Here `substochastic' means that it is positive,
i.e., $S(t):\mathcal{E}^{+}\to \mathcal{E}^{+}$, and such that
$\|S_\varepsilon (t) f\| \leq \|f\|$, holding for all $t>0$.
According to \cite{TV}, $L$ as given in (\ref{M6}) is the generator
of a positive semigroup $S= \{S (t)\}_{t\geq 0}$ that is obtained
from $S_\varepsilon$ in the limit $\varepsilon \to 0$. Then the
unique solution of (\ref{Ma}) is obtained in the form
\begin{equation}
  \label{norm60}
f_t = S(t) f_0.
\end{equation}
However, the limiting semigroup $S$ may be only substochastic -- not
stochastic, and hence the evolution $f_0 \to f_t = S(t) f_0$ may be
dishonest. The proof of its honesty -- based on the condition
$m>m_0$, see (\ref{M27z}) -- is then conducted by means of a result
of \cite{Mustapha}. The proof of the second part is conducted with
the help of methods of \cite{TV} by which we show that the semigroup
$S$ preserves the norm of $f$ defined in (\ref{norm20}) whenever
$m>m_1$. That is, under the latter condition one has
\[
\|S(t) f_0\|_{\alpha, \sigma} \leq \| f_0\|_{\alpha, \sigma},
\]
holding for all $t>0$ and some $\alpha$ and $\sigma$. By
(\ref{norm60}) this yields the property in question.

\subsection{Concluding remarks}
First of all we make some comments on the results of  Theorem
\ref{1tm}. By this statement the expected number of cells at time
$t$ is
\begin{equation*}
N(t) = \int_{\Gamma} |\gamma| f_t (\gamma) \lambda ( d \gamma),
\end{equation*}
where $|\gamma|$ stands for the number of points in $\gamma$. By
(\ref{M27}) and (\ref{norm20}) we then conclude that
\begin{equation*}
  N(t) \leq \sigma^{-1} \|f_t\|_{\alpha,\sigma}  \leq \sigma^{-1}
  \|f_0\|_{\alpha,\sigma},
\end{equation*}
holding for all $t>0$ and $m>m_1$. Then a therapeutic outcome of
Theorem \ref{1tm} is that the number of tumor cells will not
increase in time whenever the latter condition is satisfied. This
may determine the minimal level of the therapeutic pressure needed
to achieve this goal. Second,
 we note that the fulfilment of the
condition $m>m_0$ guarantees that the evolution $f_0 \to f_t$ is
honest since $f_t$ is positive and $\|f_t\|=1$. If $m_0=0$, which is
the case if $b\leq \sigma$, then the condition $m>m_1$ coincides
with that in (\ref{6}). Hence, in this case the heuristic arguments
leading to (\ref{6}) give the same answer as the microscopic
modeling resulting in Theorem \ref{1tm}. One cannot exclude,
however, that the evolution $f_0 \to f_t$ fails to be honest for
$m\in (m_*, m_0]$ if $m_0>m_*$. In this case, the individual-based
modeling yields a more precise result that is unaccessible by
heuristic theories. Note also that $b$ takes into account possible
dependence between the siblings cycle lengths, totally ignored in
the heuristic deduction of (\ref{6}).

Finally, the model defined by $L$ introduced in (\ref{M2}) can be
modified to take into account the following aspects: (a) variability
of the distribution of the lifetimes of offsprings; (b) variability
of the death rate $m$. Aspect (a) means that the density function
$G$ of a cell may be different from that of her daughters. This can
be realized by adding an additional trait $y\in Y$ with a suitable
set $Y$. The change of this trait might then be related to
mutations. Aspect (b) means dependence of $m$ on $y$ that takes into
account, e.g., drug resistance acquired in the course of mutations.
In the mathematics, introducing $y$ will correspond to passing from
single $x$ to compound traits $(x,y)$, and thus to dealing with
\emph{marked configurations}, see \cite{JK} and the papers quoted
therein. We plan to study this model in a forthcoming work.

\section*{Acknowledgments}
Yuri Kozitsky was supported  by National Science Centre, Poland
(NCN), grant 2017/25/B/ST1/00051 that is cordially acknowledged by
him.

\bibliographystyle{ws-rv-van}
\bibliography{ws-rv-sample}

\printindex                         

\end{document}